# Temporal and Spatial Aspects of Gas Release During the 2010 Apparition of Comet 103P/Hartley-2[1]

M. J. Mumma[1], B. P. Bonev[1,2], G. L. Villanueva[1,2], L. Paganini[1,9], M. A. DiSanti[1], E. L. Gibb[3], J. V. Keane[4], K. J. Meech[4], G. A. Blake[5], R. S. Ellis[6], M. Lippi[7], H. Boehnhardt[7], and K. Magee-Sauer[8]

[1]Goddard Center for Astrobiology, NASA GSFC, MS 690.3, Greenbelt, MD 20771; michael.j.mumma@nasa.gov

[2]Dept. of Physics, Catholic Univ. of America, Washington, DC 20064

[3]Department of Physics and Astronomy, University of Missouri, St. Louis, MO 63121

[4]Institute for Astronomy, Univ. of Hawaii, Honolulu, HI 96822

[5]Division of Geological & Planetary Sciences, California Institute of Technology, Pasadena, CA 91125

[6]Division of Physics, Mathematics & Astronomy, California Institute of Technology, Pasadena, CA 91125

[7]Max Planck Institute for Solar System Research, DE 37191 Katlenburg-Lindau, Germany

[8]Department of Physics and Astronomy, Rowan University, Glassboro, NJ 08028 – 1701

---

ABSTRACT

We report measurements of eight primary volatiles ($H_2O$, HCN, $CH_4$, $C_2H_6$, $CH_3OH$, $C_2H_2$, $H_2CO$, and $NH_3$) and two product species (OH and $NH_2$) in comet 103P/Hartley-2 using high dispersion infrared spectroscopy. We quantified the long- and short-term behavior of volatile release over a three-month interval that encompassed the comet's close approach to Earth, its perihelion passage, and flyby of the comet by the Deep Impact spacecraft during the EPOXI mission. We present production rates for individual species, their mixing ratios relative to water, and their spatial distributions in the coma on multiple dates. The production rates for water, ethane, HCN, and methanol vary in a manner consistent with independent measures of nucleus rotation, but mixing ratios for HCN, $C_2H_6$, & $CH_3OH$ are independent of rotational phase. Our results demonstrate that the ensemble average composition of gas released from the nucleus is well defined, and relatively constant over the three-month interval (September 18 through December 17). If individual vents vary in composition, enough diverse vents must be active simultaneously to approximate (in sum) the bulk composition of the nucleus. The released primary volatiles exhibit diverse spatial properties which favor the presence of separate polar and apolar ice phases in the nucleus, establish dust and gas release from icy clumps, and from the nucleus, and provide insights into the driver for the cyanogen (CN) polar jet. The spatial distributions of $C_2H_6$ & HCN along the near-polar jet (UT 19.5 October) and nearly orthogonal to it (UT 22.5 October) are discussed relative to the origin of CN. The ortho-para ratio (OPR) of water was 2.85 ± 0.20; the lower bound (2.65) defines $T_{spin}$ > 32 K. These values are consistent with results returned from ISO in 1997.

*Online-only material:* none

1. INTRODUCTION

Prior to its apparition in 2010, the ecliptic comet 103P/Hartley-2 (hereafter Hartley-2) was distinguished among short period comets of Kuiper Belt origin by three factors: first, its nucleus exhibited unusually high fractional activity (approaching 100% of the sunlit area; Groussin et al. 2004, Lisse et al. 2009), second, it released an unusually high (as then perceived) abundance of $CO_2$ (~10-20% relative to $H_2O$; Colangeli et al. 1999, Crovisier et al. 1999), and third, it displayed apparently 'normal' (or 'typical') composition amongst a dynamical population known for having many members depleted in (the unknown) 'long-chain' hydrocarbons whose destruction could lead to the $C_2$ and $C_3$ radicals observed in the coma (A'Hearn et al. 1995).

It is now recognized that the first two factors are connected, through $CO_2$-controlled nucleus activity that releases icy clumps into the coma; their subsequent dissolution leads to an overestimate of the fractional activity of the nucleus (A'Hearn et al. 2011). The classification as 'typical' is based on the relative production rates of reactive species ($C_2$, CN, $C_3$), which cannot be stored stably in the cometary nucleus and whose parentage is disputed or unknown. While product species are interesting, the identity and abundance of the primary volatiles and rocky grains released from the nucleus are the key clues for testing the origin and evolution of cometary material. Volatiles released directly from the nucleus (whether as gases or ices) are here termed 'primary' species, while those produced in the coma are termed 'product' species.

Production rates of primary volatiles can change dramatically as orbital motion and nucleus rotation modulate the received insolation (and thus gas production) on long and short time scales, respectively. Moreover, the apparent mixing ratios in the coma may change if nucleus regions with distinct chemical compositions are exposed to sunlight sequentially, e.g., by nucleus rotation. And, spatial differences in the release of specific volatiles can occur if individual active vents contain material of diverse volatile character. Thus, measurements acquired through snapshots in time can be placed in a global context for the nucleus only if such behavior is characterized. Unfortunately, taking such snapshots of the composition from one or two observations has long biased studies of primary volatiles in comets. To provide that contextual perspective for Hartley-2, our investigation emphasized both the long-term and short-term behavior of gas release from the cometary nucleus.

## 2. OBSERVATIONS & RESULTS

We acquired high-resolution near-infrared spectra of comet Hartley-2 using CRIRES at the ESO VLT in late July, and with NIRSPEC at the W. M. Keck Observatory from late July through mid-December 2010. The heliocentric (geocentric) distance (AU) ranged from 1.62 (0.78) on 26 July to 1.06 (0.13) on 28.25 October (perihelion), and then increased to 1.26 (0.37) on 17 December. In late July, the comet was less active than expected by about a factor of ten so we achieved upper limits only. Although the comet remained fainter than expected throughout this apparition (Combi et al. 2011, Meech et al. 2011), we successfully characterized volatile production with NIRSPEC on Sept. 18 (Mumma et al. 2010), Oct. 19 & 22, Nov. 16, and Dec. 16 & 17.

NIRSPEC spectra were acquired by nodding the telescope along the slit in an ABBA sequence, with the two beams separated by 12 arc-seconds (half the slit length). At each grating setting, spectra of infrared standard stars and calibration frames (flats and darks) were acquired for absolute flux calibration of the cometary data. We followed our standard methodology for data reduction and analysis of the individual echelle orders (Bonev 2005, DiSanti et al. 2001, Mumma et al. 2001). The signals from the two beams were combined, and spectra were extracted over nine rows (1.78 arc-sec) centered on the nucleus (taken to be the row containing the peak continuum). We isolated cometary molecular emissions by subtracting a modeled dust continuum multiplied by the atmospheric transmittance. We synthesized the spectral transmittance by using a multiple layer atmosphere and a radiative transfer model (LBLRTM) that accessed the HITRAN-2008 molecular database augmented with our custom updated line parameters and fluorescence models (e.g., Villanueva et al. 2008, Kawakita and Mumma 2011, Radeva et al. 2011, Villanueva et al. 2011).

Typical spectra are presented in Figure 1. From such spectra, we quantified $H_2O$, $CH_4$, $C_2H_2$, $C_2H_6$, $H_2CO$, $CH_3OH$, HCN, $NH_3$, $NH_2$, and OH (prompt emission) on multiple dates. A robust rotational temperature is derived from the observed water line intensities (Table 1), and the derived ortho-para abundance ratio (OPR) is 2.85 ± 0.20. The OPR standard deviation (0.20) reflects the combined effects of measurement error and modeling uncertainty for intensities of ortho and para lines. The upper bound of the OPR (3.05) is consistent with statistical equilibrium (spin temperature, $T_{spin}$ > ~ 55 K); the lower bound (2.65) defines $T_{spin}$ > 32 K. ISO observations in 1997 returned OPR values of 2.76 ± 0.08 and 2.63 ± 0.18, and $T_{spin}$ of 34 ± 3 K (Crovisier et al. 1999). For UT November 4, 2011, Dello Russo et al. (2011) reported OPR = 3.4 ± 0.3.

The long-term evolution of water production obtained from our data is compared with that derived from optical magnitudes in Figure 2a, and its short-term behavior is presented in Figure 2b. We estimated long-term production rates of water from optical magnitudes, following Jorda et al. (2008): $\log[Q(H_2O)] = \log(k) + 30.675 - 0.2453\ m_h$, where k was derived to be 0.26 and $m_h$ is the visual magnitude [$m_h$ = 8.7 + 20log $r_h(t_0 - 20)$, see Yoshida (2011)]. Our mean water production rates are in good agreement with those reported by Combi et al. (2011).

The trends in mixing ratio ($C_2H_6$, HCN, $CH_3OH$) relative to water are shown in Figure 2c. Detailed parameters for seven primary volatiles and $NH_2$ (instrument settings, rotational temperatures, production rates, and mixing ratios) on October 19 & 22 are given in Table 1 (for $CH_4$, see Table 2, footnote d). A summary of water production rates and mixing ratios for all dates is presented in Table 2. We targeted CO in mid-September and mid-December, but poor weather restricted our observations and CO was not attempted. In early November, CO was measured with HST and found to be within the range 0.15 to 0.45% (Weaver et al. 2010, Weaver et al. 2011).

The spatial profiles measured along slit are shown for $H_2O$, $C_2H_6$, and dust on four dates, and for HCN and $CH_3OH$ on two (Figure 3). The ends of the slit are marked (+, -) (see inset diagram in each panel), and corresponding pixels are so-marked on the abscissa (panels *a - d*). The '+' direction corresponds to the projection (on the sky plane) of the true comet-Sun direction, but pixels in the '+' direction sample material in both sunward and anti-sunward hemispheres (compare diagrams above each panel). For isotropic outflow, about ¾ of the sampled coma gas would be in the sunward hemisphere on UT Sept. 18, however, no exact conclusion can be reached when outflow is anisotropic. We loosely term the '+' direction as 'sampling the sunward hemisphere', and the '−' direction as 'sampling the anti-solar hemisphere'.

On UT 18 September, we aligned the slit nearly along the sun-comet line (within ~6 degrees). The spatial profile of the continuum is symmetric, but $H_2O$ is extended in the anti-solar hemisphere ('−' direction) while $C_2H_6$ is extended in the sunward hemisphere ('+' direction) (Fig. 3a).

On UT 19 October, a strong CN jet was reported with position angle nearly due North from the optocenter (Knight & Schleicher 2010, Schleicher & Knight 2011). We aligned the slit along the jet, and nearly orthogonal to the sun-comet line (Fig. 3b). The spatial profiles of continuum, $H_2O$, and $CH_3OH$ are symmetric but gases are more extended than dust in both directions. $C_2H_6$

is more extended in the jet direction ('+') than are other gases and dust, and HCN is more extended than $C_2H_6$ in the anti-jet direction (but less so than $H_2O$ and $CH_3OH$).

On UT 22 October, we aligned the slit nearly along the sun-comet line (orthogonal to the CN jet) (Fig. 3c). All gases are more extended in the sunward hemisphere ('+') than is dust and their profiles have similar shape. Water is asymmetric and more extended in the anti-solar hemisphere ('−'), compared with $CH_3OH$, HCN, $C_2H_6$ and dust.

On UT 16 December, we aligned the slit ~40 degrees to the sun-comet line. Ethane is more extended in the sunward hemisphere ('+') than is $H_2O$ or dust; $H_2O$ is more extended than dust and ethane in the anti-solar hemisphere ('−') (Fig. 3d).

3. DISCUSSION

The activity of Hartley-2 varied strongly with time. The short-term behavior is especially significant – the global production rates for water (the dominant volatile) changed by approximately a factor of two from 19 to 22 October, and again from 16 to 17 December (Figure 2b, Table 1). The production rates for water, ethane, HCN, and methanol vary in a manner consistent with nucleus rotation, similar to the behavior reported for the optical brightness of gas and dust in the coma and for radar images (Harmon et al. 2010, Harmon et al. 2011), CN jet activity (Jehin et al. 2010), activity in HCN, and other phenomena linked directly to the nucleus (Meech et al. 2011). Biver et al. (2010) reported variations in HCN with amplitude ~2.1 ($Q_{max}/Q_{min}$), with maxima observed around 25.10 and 27.24 UT and (broader) minima around ~25.4, 26.2 and ~26.9 October. Compared with our measurements, smaller amplitude is expected for the larger beam sizes used by Biver et al., because the time since release increases linearly with nucleocentric distance (assuming constant outflow velocity).

The mixing ratios of trace gases measured by us (the ratios of production rates) are steady throughout the apparition (Figure 2c), demonstrating that Hartley-2 releases material that is rather homogeneous in the bulk average. Spacecraft images of the nucleus revealed many individual vents and jets (A'Hearn et al. 2011). Production rates from individual vents may turn on and off as they rotate into and out of sunlight, but our data show little change – if any – in the mean chemical composition of released material. $CO_2$ comprised about 20% of the primary volatiles detected by the on-board spectrometer, and its sublimation is thought to control the release of 'clumps' of water ice detected in images (A'Hearn et al. 2011). If so, this apolar ice ($CO_2$) must be largely segregated from the polar water ice in the active nucleus regions. Moreover, ethane and water show distinct spatial profiles (e.g., Figure 3b) but the ethane mixing ratio ($C_2H_6/H_2O$) changes little as the nucleus rotates, suggesting that they are segregated into two distinct ice phases, as well. If ethane and carbon dioxide reside in a common apolar ice phase, carbon dioxide should show a similar steady mixing ratio (relative to co-released water) with changing rotational phase. But if it does not, then two distinct apolar ice moieties are implied along with a third polar ice. Future spacecraft results (of $CO_2/H_2O$ vs. rotational phase) may test this hypothesis.

The mean mixing ratio for HCN in Hartley-2 agrees well with those measured in several Oort cloud comets tentatively called "organics-normal" based on primary volatiles in the relatively small number (~20) of comets measured in the IR. However, the mixing ratios for ethane and methanol are slightly higher (and the mixing ratios of $C_2H_2$ and CO are much lower) in Hartley-2 than in this hypothesized "organics normal" group (Table 2) (Mumma et al. 2003, DiSanti & Mumma 2008). $CO_2$ (currently estimated at ~ 20%) is at the high end of the range found by the *Akari* space observatory for comets within 2.5 AU of the Sun (H. Kawakita, private communication; Ootsubo et al. 2010). The high abundance ratio of $CO_2/H_2O$ is not expected if hydrogenation of $C_2H_2$ and CO is responsible for the high ethane/acetylene and $CH_3OH$/CO ratios in Hartley-2 (Mumma et al. 1996). A distinct additional mechanism is needed for efficient formation of $CO_2$, perhaps by quantum-mechanical tunneling (of CO and $O(^3P)$) on CO-enriched pre-cometary grains or by prompt recombination of CO with $O(^1D)$ produced by radiation processing of icy grain mantles.

### 3.1 Interpreting Differences in Spatial Profiles.

Our data reveal distinctly different outflow properties for individual species, supporting local variations in mixing ratios (Figure 3). The extent of our sampling (along the slit) at the comet varies (linearly) with geocentric distance, e.g., on Oct. 19 and 22 the extent was ± 300 km about the nucleus, but was ± 600 km on Sept. 19 and ± 1000 km on Dec. 16 (Figure 3). On Sept. 18, Oct. 22 and Dec. 16, water is extended in the anti-solar hemisphere while ethane is extended in the sunward hemisphere (Figs. 3a, 3c, 3d). The enhancement of water in the anti-sunward direction could be associated with radiation pressure on small icy grains in the near-nucleus coma, before vaporization (but see the discussion on dissolution of 'clumps', below). As a hypervolatile, ethane is likely released directly into the coma from active regions of the nucleus in response to local insolation, and thus flows mainly into the sunward hemisphere.

On Oct. 19 we aligned the slit along the CN North polar jet (Knight & Schleicher 2010, Schleicher & Knight 2011), and nearly orthogonal to the sun-comet line (Figure 3*b*). The measured water and methanol profiles are symmetric along the slit while ethane and HCN are extended in the direction of the jet. The ethane spatial profile is congruent with the dust continuum in the anti-jet direction but is much more extended in the jet direction (where it is also relatively more extended than the other primary volatiles). This congruence is also seen in Figures 3*a, 3c,* and 3*d*. A comparison of the geometries demonstrates that the congruence is seen in the '−' direction (anti-solar) in all cases. We note that the FWHM of the continuum is similar to that of the PSF measured on the calibration stars.

The combined behavior suggests strongly that ethane is escaping directly from the sunlit nucleus surface itself, while most other primary volatiles are released by vaporization of icy aggregates ('clumps') dragged outward by escaping $CO_2$. The very slow radial velocities of escaping icy aggregates (typically 0.3 m s$^{-1}$, A'Hearn et al. 2011) imparts similar center-of-mass motion to the sublimating volatiles, which then adopt a nearly isotropic distribution determined by sublimation-endowed thermal velocities (several hundred m s$^{-1}$). The released grains will be only poorly accelerated by the subliming gas, owing to low gas densities near the sublimating clump, and will reach much smaller terminal velocities compared with grains released at the nucleus of a normal comet. This is likely a principal factor in establishing the observed rapid fall-off in the continuum profiles. At this early stage we cannot quantify the relative contributions of the nucleus and the dissolution products to the continuum detected by us, and we defer further discussion to a later publication.

The fact that HCN is much more extended in the jet direction than in the anti-jet direction is consistent with HCN outflow in the jet itself. CN is a known photolysis product of HCN, so a portion of CN in the jet is certainly produced from HCN. However, it is uncertain whether HCN is the sole source of CN in the jet, or indeed elsewhere in the coma. The strong enhancement of the apolar volatile ethane in the jet direction (Figure 3*b*) suggests that apolar volatiles (e.g., $CO_2$) could be driving the jet activity, and that other (proposed) apolar volatile precursors of CN (such as di-cyanogen, $C_2N_2$) could be enhanced there as well.

*3.2* Constancy of Mixing Ratios.

Short-term temporal variations in production rates of primary volatiles (associated with nucleus rotation) have until now been quantified in only two comets, both from the Oort cloud reservoir. Biver et al. (2009) reported 40% periodic variation in the water production rate in C/2001 Q4, with a period of 19.58 ± 0.1 h, but no other primary volatiles were so characterized. Production rates for $H_2O$, CO, $H_2CO$, and $CH_3OH$ in C/2002 T7 (LINEAR) varied with a period of 2.32 days, but their abundance ratios were constant within measurement uncertainties (Anderson 2010).

Hartley-2 is the third comet for which periodic variation in production of primary volatiles has been demonstrated, and the first for which unambiguous association with nucleus rotation can be made through imaging (A'Hearn et al. 2011, Harmon et al. 2010, Harmon et al. 2011). Among comets from the Kuiper Belt, Hartley-2 is thus unique in being so categorized. Considering the wide array of information gleaned from the flyby and from supporting investigations, Hartley-2 joins 1P/Halley (an Oort cloud comet) in being uniquely characterized amongst comets from the two principal reservoirs. Their similarity in composition is remarkable for implying a common heritage for (some) icy bodies in these two disparate reservoirs.

Keck telescope time was granted by NOAO (through the Telescope System Instrumentation Program funded by NSF), the University of Hawaii, and the California Institute of Technology. VLT time was granted by the European Southern Observatory. We gratefully acknowledge support by the NSF Astronomy and Astrophysics Research Grants Program (PI/co-PI Bonev/Gibb), by the NASA Astrobiology Institute (PI Meech, PI Mumma), by NASA's Planetary Astronomy (PI DiSanti, PI Mumma), Planetary Atmospheres (PI DiSanti; PI Villanueva), and Discovery (Meech) Programs, and by the German-Israel Foundation (D. Prialnik; M. Lippi).

References

A'Hearn, M. F., Millis, R. L., Schleicher, D. G., Osip, D. J., & Birch, P. V. 1995, Icarus, 118, 223

A'Hearn, M. F., et al. 2011, Science (in press)

Anderson, W. M. 2010, Ph. D. thesis, Catholic University of America

Biver, N., Biver, N., Bockelée-Morvan, D., Colom, P., Crovisier, J., & Lecacheux, A. 2009, A&A 501, 359

Biver, N., et al. 2010 (private communication)

Bonev, B. P. 2005. Ph.D. thesis, Univ. Toledo, http://astrobiology.gsfc.nasa.gov/Bonev_thesis.pdf

Colangeli, L., et al. 1999, A&A, 343, L87

Combi, M. R., Bertaux, J.-L., Quémarais, E., Ferron, S., Mäkinen, J. T. T. 2011, Ap. J., 732:Lxxx (this issue, in press)

Crovisier, J., et al. 1999, In *The Universe as Seen by ISO*, ed. P Cox, MF Kessler, ESA-SP 427:161. Noordwijk, Neth.: ESA

Dello Russo, N., et al. 2011, Ap. J., 732:Lxxx (this issue, in press)

DiSanti, M. A. & Mumma, M. J. 2008, Space Sci. Rev., 138, 127

DiSanti, M. A., Mumma, M. J., Dello Russo, N., & Magee-Sauer, K. 2001, Icarus, 153, 361

Groussin, O., Lamy, P., Jorda, L., & Toth, I. 2004, A&A, 419, 375

Harmon, J. K., Nolan, M. C., Howell, E. S., & Giorgini, J. D. 2010, IAU Circ., 9179

Harmon, J. K., Nolan, M. C., Howell, E. S., & Giorgini, J. D. 2011, Ap. J., 732:Lxxx (this issue, in press)

Jehin, E., Manfroid, J., Hutsemekers, D., Gillon, M., & Magain, P. 2010, CBET, 2589

Jorda, L., Crovisier, J., & Green, D. W. E. 2008, in Asteroids, Comets, Meteors 2008, LPI Contrib. 1405, 8046

Kawakita, H. & Mumma, M. J. 2011, Ap. J. 727, 91

Knight, M., & Schleicher D. 2010, IAU Circ. 9175

Knight, M. M., & Schleicher, D. G. 2011, AJ (in press, arXiv:1103.5466)

Lisse, C., et al. 2009, PASP, 121, 968

Meech, K. J., et al. 2011, Ap. J., 732:Lxxx (this issue, in press)

Mumma M. J., et al. 2001, Science, 292, 1334

Mumma, M. J., DiSanti, M. A., Dello Russo, N., Magee-Sauer, K., Gibb, E. L., & Novak, R. 2003, Adv. Space Res., 31, 2563

Mumma, M. J., et al. 1996, Science, 272, 1310

Mumma, M. J., et al. 2010, IAU Circ. 9180; corrigendum, *ibid.* 9186

Ootsubo T, et al. 2010. Ap. J., 717:L66

Radeva, Y. L., Mumma, M. J., Villanueva, G. L., & A'Hearn, M. F. 2011 ApJ, 729, 135

Schleicher, D. G. & Knight, M. 2011, AJ, in press

Villanueva, G., Mumma, M. J., Novak, R., & Hewagama, T. 2008, Icarus, 195, 34

Villanueva, G., Mumma, M. J., & Magee-Sauer, K. 2011, JGR (submitted)

Weaver, H. A., Feldman, P. D., A'Hearn, M. F., Dello Russo, N., & Stern, S. A. 2010, IAU Circ. 9183

Weaver, H. A., Feldman, P. D., A'Hearn, M. F., Dello Russo, N., & Stern, S. A. 2011, Ap. J., 732:Lxxx (this issue, in press)

Yoshida, S. 2011, http://www.aerith.net

# Table 1. Molecular Parameters for Primary Volatiles in 103P/Hartley-2

**UT 19 October**

| Species | Time | Setting/order [a] | $T_{rot}$ °K | NC Production Rate [b] Molecules/sec | GF | Global Q [c] Molecules/sec | Abundance Ratio % |
|---|---|---|---|---|---|---|---|
| $H_2O$ | 11:21 – 12:20 | KL2/26 | [d]86 ± 4 | 242 ± 12.7 E25 | 1.7 | 411 ± 21.5 E25 | 100 |
| HCN | " | KL2/25 | [d]76 ± 6 | 52.8 ± 1.87 E23 | " | 89.8 ± 3.18 E23 | 0.22 ± 0.01 |
| $C_2H_2$ | " | " | (80) | 27.6 ± 3.90 E23 | " | 47.0 ± 6.63 E23 | 0.11 ± 0.02 |
| $NH_3$ | " | " | (80) | 135 ± 35.3 E23 | " | 230 ± 60.1 E23 | 0.56 ± 0.15 |
| $NH_2$ [e] | " | " | (80) | 43.2 ± 7.34 E23 | " | 73.5 ± 12.5 E23 | 0.18 ± 0.03 |
| $CH_3OH$ | " | KL2/22 | (80) | 687 ± 115 E23 | " | 1168 ± 196 E23 | 2.84 ± 0.50 |
| $H_2CO$ | | KL2/21 | (80) | 80.2 ± 23.1 E23 | " | 136 ± 39 E23 | 0.33 ± 0.10 |
| $H_2O$ | 15:04 – 15:35 | KL1/26 & 27 | [d]80 ± 2 | 196 ± 2.31 E25 | 1.75 | 343 ± 4.05 E25 | 100 |
| $C_2H_6$ | " | KL1/23 | [d]73 +8/-10 | 147 ± 4.6 E23 | " | 257 ± 8.1 E23 | 0.75 ± 0.03 |
| $CH_3OH$ | " | KL1/22 | (80) | 619 ± 75.1 E23 | " | 1083 ± 131 E23 | 3.15 ± 0.38 |

**UT 22 October**

| Species | Time | Setting/order [a] | $T_{rot}$ °K | NC Production Rate [b] Molecules/sec | GF | Global Q [c] Molecules/sec | Abundance Ratio % |
|---|---|---|---|---|---|---|---|
| $H_2O$ | 11:44 – 12:42 | KL2/26 | [d]80 +1/-2 | [f]399 ± 15.0 E25 | 1.7 | 678 ± 25 E25 | 100 |
| HCN | " | KL2/25 | [d]75 +5/-6 | 105 ± 4.52 E23 | " | 179 ± 7.68 E23 | 0.26 ± 0.02 |
| $C_2H_2$ | " | " | (80) | 32.7 ± 5.22 E23 | " | 55.5 ± 8.87 E23 | 0.08 ± 0.01 |
| $NH_3$ | " | " | (80) | 177 ± 37.5 E23 | " | 301 ± 63.7 E23 | 0.44 ± 0.10 |
| $NH_2$ [e] | " | " | (80) | 54.1 ± 11.2 E23 | " | 92.0 ± 19.0 E23 | 0.14 ± 0.03 |
| $C_2H_6$ | " | KL2/22 | (80) | 300 ± 25.7 E23 | " | 509 ± 43.7 E23 | 0.75 ±0.07 |
| $CH_3OH$ | " | " | (80) | 910 ± 149 E23 | " | 1547 ± 254 E23 | 2.28 ± 0.38 |
| $H_2CO$ | " | KL2/21 | (80) | 92.6 ± 21.3 E23 | " | 157 ± 36.1 E23 | 0.23 ± 0.05 |

[a] NIRSPEC instrument setting. [b] Nucleus Centered (NC). [c] Global production rate after applying a measured growth factor (GF) to the NC production rate. [d] We tabulate the retrieved $T_{rot}$ and confidence limits; however, all measured temperatures are consistent with 80K on 19 & 22 October. For HCN and $C_2H_6$, the derived Q's are only weakly sensitive to $T_{rot}$, so we adopted 80K for all species when calculating the NC production rates. [e] $NH_2$ is a photolysis product and not a primary volatile. For a primary volatile, the inscribed sphere of the nucleus-centered pencil beam contains a fraction (~2/π) of the total beam content (for uniform spherical outflow). If a product of $NH_3$, $NH_2$ is severely depleted within the inscribed sphere (cf. Kawakita & Mumma 2011) making the apparent production rates for $NH_2$ given here a lower bound to those of $NH_3$, which likely are larger by about a factor of three. Although qualitatively consistent with the production rate measured for $NH_3$ itself, we defer further analysis of $NH_2$ to a later publication. [f] The reported error in production rate includes line-by-line scatter in measured column densities, photon noise, systematic uncertainty in the removal of the cometary continuum, and (minor) rotational temperature uncertainty.

## Table 2. Summary of Abundance Ratios

| Date[a], UT | Q(H$_2$O), 10$^{25}$ s$^{-1}$ H$_2$O | Abundance Ratio[b],% | | | | | |
|---|---|---|---|---|---|---|---|
| | | C$_2$H$_6$ | C$_2$H$_2$ | CH$_3$OH | HCN | NH$_3$ | H$_2$CO |
| 26 July[c] | < 195[e] | < 1.5[e] | — | — | — | — | — |
| 18 Sept[d] | 213 ± 15 | 0.76 ± 0.09 | — | 2.61 ± 0.56 | — | — | — |
| | 314 ± 59 | — | < 0.13[e] | — | 0.29 ± 0.06 | < 0.93[e] | — |
| 19 Oct | 411 ± 22 | — | 0.11 ± 0.02 | 2.84 ± 0.50 | 0.22 ± 0.01 | 0.56 ± 0.15 | 0.33 ± 0.10 |
| | 343 ± 4.1 | 0.75 ± 0.03 | — | 3.15 ± 0.38 | — | — | — |
| 22 Oct | 678 ± 26 | 0.75 ± 0.07 | 0.08 ± 0.01 | 2.28 ± 0.38 | 0.26 ± 0.02 | 0.44 ± 0.10 | 0.23 ± 0.05 |
| 16 Nov | 766 ± 75 | 0.69 ± 0.12 | 0.09 ± 0.03 | 2.60 ± 0.68 | 0.23 ± 0.03 | 0.94 ± 0.20 | — |
| 16 Dec | 662 ± 31 | 0.64 ± 0.06 | — | 3.00 ± 0.42 | — | — | — |
| 17 Dec | 296 ± 71 | < 0.81[e] | — | — | 0.23 ± 0.08 | — | — |
| "Organics Normal"[f] | | 0.59 ± 0.03 | 0.24 ± 0.03 | 2.2 ± 0.2 | 0.26 ± 0.03 | — | N/A |

[a]The heliocentric distance (AU), geocentric distance (AU), and geocentric velocity (km s$^{-1}$) were: 18 Sept (1.191, 0.262, -11.9), 19 Oct (1.065, 0.121, -0.7), 22 Oct (1.062, 0.121, +0.4), 16 Nov (1.092, 0.212, +0.9), 16 Dec (1.25, 0.371, +9.9), 17 Dec (1.26, 0.371, +9.9), respectively.

[b]Mixing Ratios are expressed relative to water (see Table 1). To derive individual production rates, we applied fluorescence models based on these adopted rotational temperatures: 50 K for 26 July; 65 K for 18 Sept. and 16 & 17 Dec.; 80 K for 19 & 22 Oct. (see Table 1); and 75 K for Nov. 16. Our mixing ratios agree well with those of Dello Russo et al. (2011), excepting methanol. We extracted production rates from CH$_3$OH $\nu_3$ Q-branch intensities (KL1/22), using g-factors (at 1 AU heliocentric distance) equal to 1.13 x 10$^{-5}$ for 18 Sept. and 16 Dec., and 1.02 x 10$^{-5}$ for 19 October. The g-factor used by Dello Russo et al. (2011) is larger by about a factor of two, leading to correspondingly smaller production rates and mixing ratios. On 19, 22 Oct., and 16 Nov., we employed an empirical g-factor ([1.87 ± 0.31] x 10$^{-5}$) for lines of the $\nu_2$ and $\nu_9$ bands in the range 2919 – 2929 cm-1 (KL2/22) and uncertainties in mixing ratios for CH$_3$OH on these dates include this uncertainty in g-factor.

[c]On July 26, we obtained upper limits (3-$\sigma$) for production rates (Q) of three primary species. Using CRIRES they were: Q(C$_2$H$_6$) < 3.0 x 10$^{25}$ s$^{-1}$ and Q(CH$_4$) < 11 x 10$^{25}$ s$^{-1}$. Using NIRSPEC: Q(H$_2$O) < 195 x 10$^{25}$ s$^{-1}$.

[d]We searched for methane but did not detect it. The upper limit (3-$\sigma$) to its abundance ratio was 3% on 18 September and 1.2% on 16 November.

[e]Upper limits are at the 3-$\sigma$ confidence level.

[f]This compositional grouping is dominant among Oort cloud comets, in a taxonomy based on primary volatiles (Mumma et al. 2003, DiSanti and Mumma 2008). The CO abundance ratio ranged from 1.8 to 15% in this group, but in Hartley-2 CO was unusually low, with an abundance ratio 0.15 to 0.45% relative to water (Weaver et al. 2010; 2011).

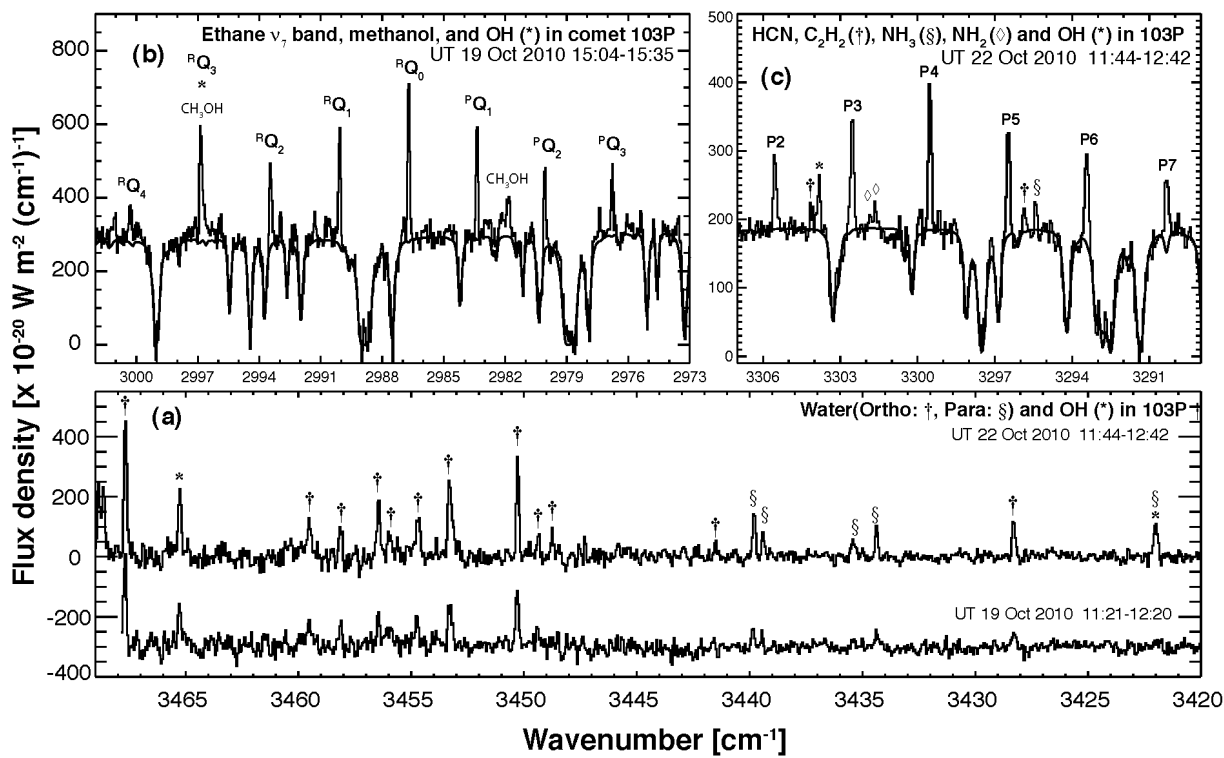

**Figure 1.** Nucleus-centered spectra of water, ethane, methanol, and hydrogen cyanide in Hartley-2, acquired in mid-October 2010. Spectra on other dates cover similar frequency ranges. **(a).** On UT 19.5 and 22.5 October, multiple spectral lines of ortho- (†) and para- (§) water are seen, along with OH prompt emission (*). A robust rotational temperature is derived from the observed water line intensities (Table 1), and the derived ortho-para abundance ratio (OPR) is 2.85 ± 0.20 (see text). The OPR standard deviation (0.20) reflects the combined effects of measurement error and modeling uncertainty for intensities of ortho and para lines. The upper bound of the OPR (3.05) is consistent with statistical equilibrium (spin temperature, $T_{spin} > \sim 55$ K); the lower bound (2.65) defines $T_{spin} > 32$ K. ISO observations in 1997 returned OPR values of 2.76 ± 0.08 and 2.63 ± 0.18, and $T_{spin}$ of 34 ± 3 K (Crovisier et al. 1999). For UT November 4, 2011, Dello Russo et al. (2011) reported OPR = 3.4 ± 0.6. **(a).** The spectrum of ethane and methanol on 19.5 October. The rotational temperature derived for ethane agrees with that of co-measured water on that date (panel *a*, Table 1). **(c).** The spectrum of HCN (P-lines), $C_2H_2$ (†), $NH_3$ (§), $NH_2$ (◊), and OH (*) on UT 22.5 October. The rotational temperature derived for HCN agrees with that of co-measured water on that date (panel *a*, Table 1).

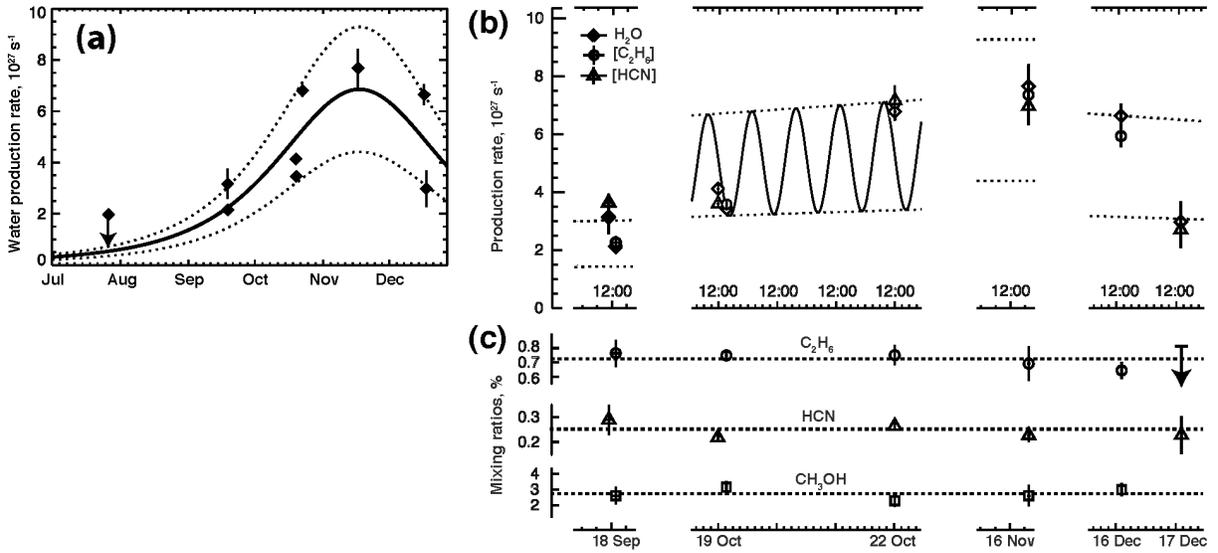

**Figure 2.** Comparison of our measured water production rates and estimated rates derived from visual magnitudes over the course of our campaign (see §2). **(a).** The measured and estimated production rates are closely correlated over the long-term. The dashed lines embrace the amplitude excursions ($Q_{max}/Q_{min}$ = 2.1) in the short-term production rates that we observe, probably the signature of nucleus rotation. **(b).** Short term variations in water production. The measurements of UT 19 and 22 Oct are compared with a rotational phase curve with an assumed period of 18 hours, an excursion amplitude of 2.1 and time phase consistent with the times of maxima and minima reported from radio observations (Biver et al. 2010). Dotted lines mark the high and low limits of production, as in panels *a* and *c*. **(c).** Mixing ratios (relative to water), for $C_2H_6$, HCN, and $CH_3OH$. Their constancy contrasts starkly with the strong variation in individual production rates, and demonstrates the bulk homogeneity of these primary volatiles in the cometary nucleus.

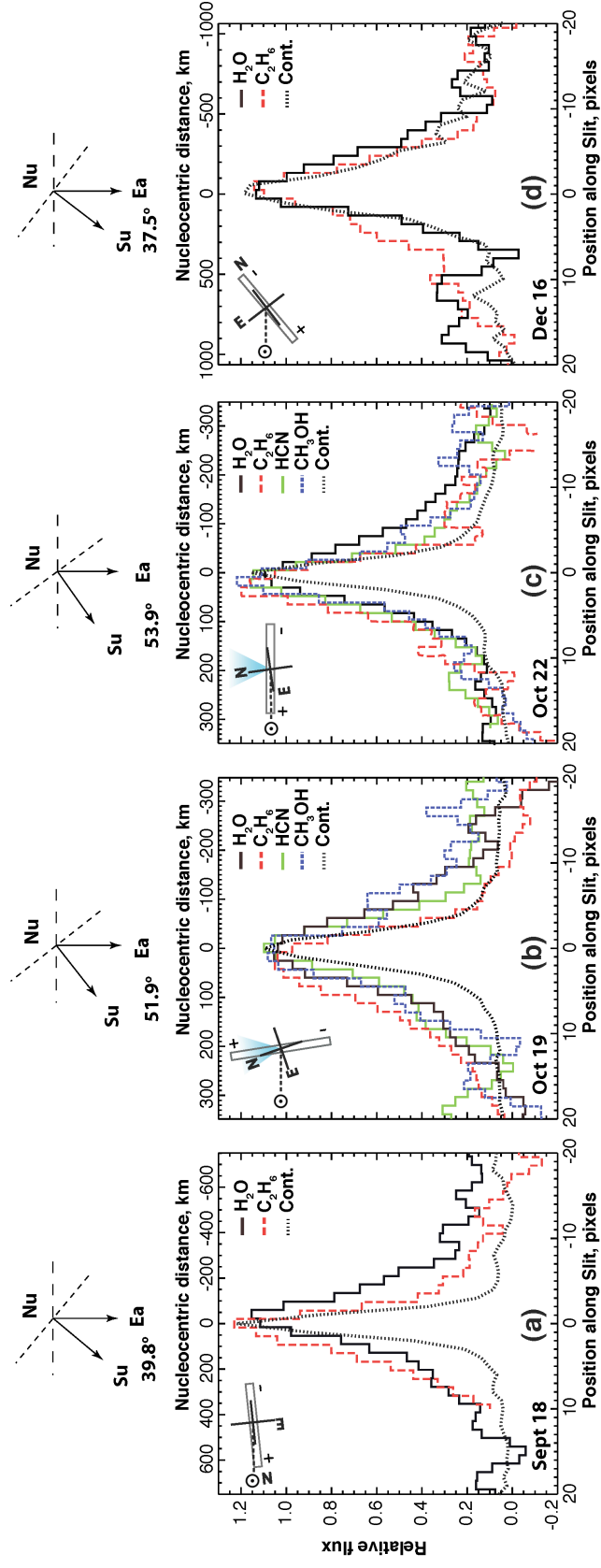

**Figure 3.** Spatial profiles of primary volatiles in the sky plane, measured during the 2010 apparition of Hartley-2. In each panel (*a - d*), the Sun-comet line is horizontal, and the cardinal directions and slit orientation are shown (insets). Spatial scales at the comet change with date, and are displayed at the top of each panel. The view looking down on the Sun-Comet-Earth plane is shown above each panel ("Su", "Nu", and "Ea" denote positions of the Sun, comet nucleus (at the vertex), and Earth). The plane dividing sunward and antisunward hemispheres is marked (short dashes), as is the sky plane at the comet (long dashes). **(a).** UT 18 September. We aligned the slit nearly along the Sun-comet line (within ~ 6 degrees). **(b).** UT 19 October. A CN jet was reported with position angle nearly due North (Knight & Schleicher 2010). We aligned the slit along the jet, and nearly orthogonal to the Sun-comet line. **(c).** UT 22 October. We aligned the slit nearly along the Sun-comet line (orthogonal to the CN jet). **(d).** UT 16 December. We aligned the slit ~40 degrees to the Sun-comet line. Spatial profiles are described in §2 and discussed in §3.1.